\documentclass[runningheads]{svmult}
\usepackage{epsfig,multicol,amsmath,amssymb,setspace,graphicx,subfigure,axodraw}
\usepackage{makeidx}   
\usepackage{graphicx}  
\usepackage{multicol}  
\usepackage{cropmark} 
\usepackage{physprbb}  
\begin{document}
\title*{QCD Axion and Quintessential Axion\footnote{Talk presented
at {\it Beyond'03}, Castle Ringberg, Germany, June 9--14, 2003.}}
\toctitle{Focusing of a Parallel Beam to Form a Point
\protect\newline in the Particle Deflection Plane}
%
%
\titlerunning{Axions}
\author{Jihn E. Kim}
\authorrunning{Jihn E. Kim}
\institute{School of Physics, Seoul National University, Seoul
151-747, Korea }

\maketitle

\begin{abstract}
The axion solution of the strong CP problem is reviewed together
with the other strong CP solutions. We also point out the
quintessential axion(quintaxion) whose potential can be extremely
flat due to the tiny ratio of the hidden sector quark mass and
the intermediate hidden sector scale. The quintaxion candidates
are supposed to be the string theory axions, 
the model independent or the model dependent axions.
\end{abstract}

\def\thetab{$\bar\theta$}
\def\theq{$\theta_{QCD}$}
\def\thew{$\theta_{weak}$}
\section{The Strong CP Problem}
At this conference, the organizer asked to review on
the axion~\cite{axionr} as the CDM candidate~\cite{pww}. 
Therefore, I will review the QCD axion and the related
problem, the strong CP problem. 
With supersymmetrization, an O(GeV) axino can be a CDM
candidate also~\cite{CDMaxino} in addition to the
axion CDM, but we will not discuss this interesting
class of CDM candidate here. It was reviewed
in another DM conference~\cite{Cape}. Out of several 
CDM candidates, the axion is most
attractive since it arises from the need to solve the strong CP
problem. One crucial difference of the axion CDM from most other
CDM candidates is that the axion is a classical 
field oscillation~\cite{pww}
while many other CDM candidates rely on their 
heavy particle nature.
The classical axion potential is extremely flat as 
shown in Fig. 1.
The axion value $\langle a\rangle$ stays at some point(the bullet)
for a long time before it starts to oscillates, which occurs at
$T\simeq $ 1~GeV when the Hubble time $1/H$ becomes larger than
the oscillation period $m_a$.

\begin{figure}
\begin{center}
\begin{picture}(380,40)(40,0)
\Line(50,10)(330,10) \CArc(200,600)(590,256.5,283)
\Text(330,10)[c]{$\blacktriangleright$} \Text(355,10)[c]{\Large$\
< a >$} \Text(200,15)[c]{\Large$\blacktriangledown$}
\CCirc(310,25){4}{0}{0}
\end{picture}
\caption{The very flat axion potential. The minimum of the
potential is shown by a triangle. The initial axion vacuum is
shown as a bullet.}
\end{center}
\end{figure}
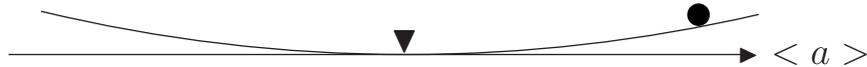
\vskip -0.5cm

This axion is the invention from the need to solve the strong 
CP problem. The instanton solution in nonabelian gauge theories
needs an effective term parametrized by~\cite{cdg} 
\begin{equation}
\frac{\bar\theta}{32\pi^2}F\tilde F
\end{equation}
where $F\tilde F\equiv \frac12\epsilon_{\mu\nu\rho\sigma}
F^{a\mu\nu}F^{a\rho\sigma}$ is the Pontryagin term of the
gluon field strength $F^a_{\mu\nu}$. The \thetab\  parameter
is the final low energy value, taking into account the weak
CP violation. It can be split into two pieces \thetab = \theq
+ \thew\ where \theq\ is the parameter determined at
high energy scale and \thew$={\rm Arg. Det.} M_q$ is the one 
generated at the electroweak scale. Since \thetab\ contributes
to the neutron electric dipole moment(NEDM), the experimental
upper bound on NEDM constrains $|\bar\theta|<10^{-9}$, which is a
fine-tuning problem: the so-called {\it strong CP problem}. 

One may argue that there is no strong CP problem 
from the beginning. One such argument is a 5D extension where
there is no 5D instanton solution~\cite{5D}. However, I think
that in the 4D effective theory language where 5D is hidden
one should consider the \thetab$F\tilde F$ term
at low energy, and the strong CP problem reappears.

So, in this talk I will review the solutions of the strong CP
problem first on, 
\begin{itemize}
\item
the calculable \thew,
\item
the massless up quark solution, and
\item
the very light axion,
\end{itemize}
and then go on to discuss an axion-like particle,
\begin{itemize}
\item
the quintessential axion~\cite{quintaxion}. 
\end{itemize}

The calculable \thew\ and the massless up quark cases
are discussed in the subsequent subsections, and 
the very light axion and quintessential axion in
separate sections.

\subsection{Calculable \thew}
In this kind of theories \theq\ is supposed to be zero,
which can be implemented by assuming the CP invariance of
the Lagrangian. Thus, the weak CP violation can be 
introduced only spontaneously~\cite{TDLee}.
After the weak CP violation is introduced spontaneously,
\thew\ can be calculated. In the beginning, this kind of
models were proposed by changing the weak 
interactions~\cite{Beg}. However, the Kobayashi-Maskawa(KM)
type weak CP violation seems to be working, and the
above spontaneous weak CP violations are not working. In
this regards, the Nelson-Barr approach~\cite{NB} of 
the spontaneous
CP violation at a very high energy scale with a calculable
\thew\  has attracted
a great deal of attention since at low energy it leads to 
the KM type weak CP violation. However, it needs superheavy
particles with a vectorlike representation at high
energy scale.  

\subsection{Massless up quark}
Suppose that we chiral-transform a quark field $q$,
\begin{equation}\label{chiral}
q\longrightarrow e^{i\gamma_5\alpha}q
\end{equation}
It is equivalent to changing \thetab$\rightarrow$\thetab$-
2\alpha$. Thus, if a theory admits such a transformation,
i.e. if there exists a massless quark,
then the strong CP problem is not present. 
The up quark is most promising for the massless
quark possibility. It was known from the very early
days of the \thetab\ vacuum that a massless up quark
solves the strong CP problem. The current 
debate~\cite{kaplan,ckim,leut,choi} on
the massless up quark solution is a phenomenological 
one. The famous up/down ratio $Z=m_u/m_d=\frac59$~\cite{Weinberg} 
seems to rule out the $m_u=0$ possibility.

But the story below 100 GeV is not so simple. The point is
that there exists the 't Hooft determinental 
interaction~\cite{thooft} shown in Fig.~2.


\begin{figure}
\begin{center}
\begin{picture}(400,220)(30,60)
\LongArrow(120,260)(180,220)\Text(100,260)[c]{\Huge $u_{L}$}
\LongArrow(120,200)(180,200)\Text(100,200)[c]{\Huge $d_{L}$}
\LongArrow(120,140)(180,180)\Text(100,140)[c]{\Huge $s_{L}$}

\CCirc(200,200){8}{0}{0}

\LongArrow(220,220)(280,260)\Text(300,260)[c]{\Huge$ u_{R}$}
\LongArrow(220,200)(280,200)\Text(300,200)[c]{\Huge$ d_{R}$}
\LongArrow(220,180)(280,140)\Text(300,140)[c]{\Huge$ s_{R}$}

\DashCurve{(120,140)(200,100)(280,140)}{4} \Text(200,100)[c]{\Huge
$\times$}
\end{picture}
\vskip -0.8cm
\caption{The 't Hooft determinental interaction is shown as
arrows. The $s$ quark line can be closed to give an effective
instanton generated $\frac{m_s}{\Lambda}\bar u_Ru_L \bar d_Rd_L$
interaction.}
\end{center}
\end{figure}
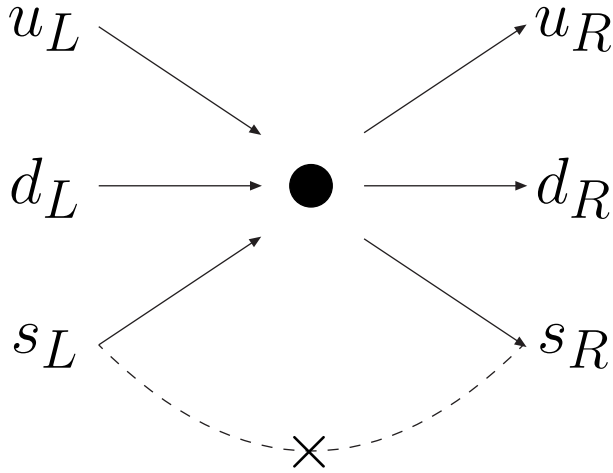

\noindent The hard quark mass considered in the chiral 
perturbation theory is hard at the 300~MeV scale. If the 
instanton generated mass is defined much above 300~MeV,
then it will act like a hard mass in the chiral perturbation
theory. Indeed, by closing the strange and down quarks\footnote{
By connecting the $u_R,s_R$ quark lines of Fig. 2 with $u_L,s_L$
lines in the $G_F$ order weak interaction, the instanton
generated $\Delta I=\frac12$ was obtained long time 
ago~\cite{kim79}.} in Fig.~2, one can obtain an up quark mass of
order
\begin{equation}
m_u=\frac{m_dm_s}{\Lambda}
\end{equation} 
where $\Lambda$ is the instanton related QCD parameter
of order GeV. Suppose that the electroweak symmetry breaking
gives a massless up quark. Then, there is no strong CP
problem. But the chiral perturbation calculation can be
performed with the above instanton generated nonzero
up quark mass. So, the massless up quark problem is
whether the instanton generated up quark mass is
consistent with the phenomenology of the chiral
perturbation. Here, there are two groups sharing the
opposite view:

Kaplan-Manohar, Choi: Yes, the up quark can be massless.

Leutwyler: No, the up quark cannot be massless.

\noindent Kaplan and Manohar~\cite{kaplan} considered
the second order chiral perturbation theory and found
that $Z=\frac{m_u}{m_d}\sim 0.2$, and concluded that
$m_u=0$ is not ruled out. In fact, they considered the
$L_7$ term in the chiral perturbation theory,
$L_7\langle \chi^\dagger U-\chi U^\dagger\rangle$ where 
$\chi$ is a $3\times 3$ mass matrix and $U$ is a $3\times 3$ 
unitary matrix in terms of the octet pseudoscalar mesons,
and obtained for the case of $m_u=0$
\begin{equation}
L_7\simeq (1\sim 2)\times 10^{-4}\ \ [\ {\rm or}\ 
\ (2L_8-L_5)_{m_u=0}\simeq -(1.2\sim 2.6)
\times 10^{-3}]
\end{equation}
where $L_8$ and $L_5$ are other terms in the chiral 
perturbation. Leutwyler~\cite{leut} attempted to 
compute $L_7$ using the QCD sum rule with the 
$\eta^\prime$(the SU(3) singlet meson) dominance,
in analogy with the vector dominance model, and obtained
$L_7\simeq -(2\sim 4)\times 10^{-4}$. If Leutwyler's
calculation is correct, the $m_u=0$ possibility is
ruled out. However, there are arguments con Leutwyler.
For example, the $\eta^\prime$ dominance can be questioned.
To answer this problem, Choi~\cite{choi} followed the
modern interpretation for the $\eta^\prime$ mass,
i.e. the instanton contribution to the $\eta^\prime$
mass, in which case he showed that he could change the
sign of $L_7$, to $L_7\simeq (3\sim 8)\times 10^{-4}$. Choi's
result is obtained by semiclassical approximation
supplemented by gluon condensate which may be the only
unreliable point. Anyway, it was noted that instanton effects
suppress $\eta^\prime$ contribution to $L_7$.
So, it is fair to say that the problem is not completely
settled in the chiral perturbation theory.

There has been a new development in the lattice calculation
of the chiral perturbation parameter. For example,
a lattice calculation~\cite{lattice} gives~\cite{ack}
\begin{equation}
2L_8-L_5\simeq (9\pm 4_{\rm stat}\pm 20_{\rm syst})\times 
10^{-5}
\end{equation} 
with $m_u/m_d\simeq 0.484\pm 0.027$. If it is true,
the $m_u=0$ possibility is closed. However, there can
be questions on the lattice calculation such as the
small number of lattice sites and no inclusion of
the instanton effects. Anyway, the possibility of
$m_u=0$ is raised from the instanton contribution
to the up quark mass. Therefore, it is fair to say that
the $m_u=0$ possibility is not closed yet from lattice
calculation.

\section{The QCD Axion}

Axion was noted~\cite{PQWW} from the spontaneously broken 
Peccei-Quinn symmetry~\cite{PQ}. More generally, we can
define an axion $a$ as the pseudoscalar WITHOUT ANY POTENTIAL 
EXCEPT that arising from the following nonrenormalizable
coupling,
\begin{equation}\label{axionanomaly}
\frac{a}{F_a}\{F\tilde F\}\equiv
\frac{1}{32\pi^2}\frac{a}{F_a}\frac12\epsilon_{\mu\nu
\rho\sigma}F^{a\mu\nu}F^{a\rho\sigma}
\end{equation} 
where $F_a$ is called $\lq\lq$axion decay constant".
As shown in Table 1, this nonrenormalizable interaction can 
arise in several different ways~\cite{witten,compax,PQ,KSVZ,DFSZ}.
\begin{table}
\caption{Nature of the axion}
\begin{center}
\renewcommand{\arraystretch}{1.4}
\setlength\tabcolsep{5pt}
\begin{tabular}{ll}
\hline\noalign{\smallskip}
Source of the non-renormalizable interaction & $F_a\sim$ \\
\noalign{\smallskip}
\hline
\noalign{\smallskip}
Higher dimensional fundamental 
interaction\cite{witten,choikim,Maxion}  
& Planck scale $M_P$ \\
In composite models\cite{compax,chun} & Compositeness scale $M_c$  
\\
Spontaneously broken PQ symmetry\cite{PQ,KSVZ,DFSZ} & PQ symmetry 
breaking scale
$\tilde v$ \\
\hline
\end{tabular}
\end{center}
\label{Tab1.1a}
\end{table}


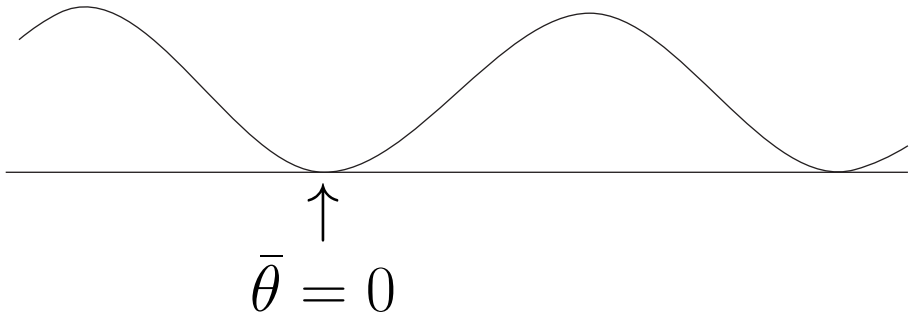
\begin{figure}
\begin{center}
\begin{picture}(400,150)(10,-20)

\Line(10,50)(350,50)
\Curve{(15,100)(30,110)(130,50)(230,110)(330,51)(350,60)}
\Text(130,35)[c]{\Huge $\uparrow$}
\Text(130,8)[c]{\Huge$\bar{\theta}=0$}
\end{picture}
\vskip -0.3cm
\caption{The minimum of the free energy of $\bar\theta$ parameter
is at $\bar\theta=0$.}
\end{center}
\end{figure}

The axion potential arising after integrating out
the gluon field has the following two properties,
\begin{itemize}
\item It is periodic with the periodicity $2\pi F_a$,\\
\item Its minimum is at $\langle a\rangle =2n\pi F_a\ 
(n={\rm integer})$~\cite{PQ,vw}.
\end{itemize}
Taking into account the above points, one can write
a cosine potential of $a$: $-\Lambda^4_{QCD}\cos(a/F_a)$.
The height of the potential is determined by the strong
interaction scale $\Lambda_{QCD}$. For the vanishing
cosmological constant, we can write the potential as
$\Lambda_{QCD}^4(1-\cos\frac{a}{F_a})$, where 
$\Lambda_{QCD}^4$ is an educated guess. Actually, 
$\Lambda_{QCD}^4$ must be qualified by considering the axion
mixing with pions. In the current algebra or in the chiral
perturbation calculation, one can fix this coefficient to 
obtain~\cite{bardeen},\footnote{Actually, the exact expression
for the axion potential from chiral
symmetry is~\cite{axionr}
$V[a]=f_\pi^2m_\pi^2\left(1-\sqrt{m_u^2+m_d^2
+2m_um_d\cos(a/F_a)/(m_u+m_d)}\right)$. 
The expansion upto the $\cos(a/F_a)$ term gives the
result (\ref{axionV}).}
\begin{equation}\label{axionV}
V[a]=\frac{Z}{(1+Z)^2}f_\pi^2 m_\pi^2\left(
1-\cos\frac{a}{F_a}\right).
\end{equation}
As we know that a massless quark makes the axion potential
flat, we can guess that the coefficient is 
$m_um_d\Lambda_{QCD}^2$(or $m_u\Lambda^3_{QCD}$ at low energy) 
which turns out to be (number)$\times
f_\pi^2 m_\pi^2$. We have to remember that the overall 
coefficient has the small quark masses, which will be very
useful in the discussion of quintaxion.
The shape of the potential is shown in Fig. 3.

It shows that at the minimum
of the axion potential \thetab\ = 0, solving the strong
CP problem. Basically, the axion solution of the strong
CP problem is a cosmological solution. Namely, in the
evolving universe, the classical axion field seeks the
minimum where
\begin{equation}
\bar\theta=0,
\end{equation}
whatever happenned before. But note that the weak interaction
violates the CP invariance and introduces a small CP odd
term in the axion potential, which shifts the \thetab\ minimum
slightly to $\bar\theta\sim 10^{-17}$~\cite{randall}. 

From (\ref{axionV}), one can easily compute the
very light axion mass
\begin{equation}
m_a=\frac{\sqrt{Z}}{1+Z}\frac{f_\pi m_{\pi^0}}{F_a}
\simeq 0.6\ [{\rm eV}] \times \frac{10^7\ {\rm GeV}}{F_a}
\end{equation}

From the early days of axion physics, the PQWW axion
seemed to be in conflict with laboratory 
experiments~\cite{Tokyo}, which was the reason that a
number of calculable \thew\ models were tried in 
1978~\cite{Beg}. From the cosmological point of view,
the PQWW axion is not of much significance since its
lifetime is around $10^{-8}$~s order. This has led to the 
invention of the invisible axions~\cite{KSVZ,DFSZ}, but it is
proper to call them as very light axions since in the
Sikivie type cavity experiments~\cite{cavity} one may 
detect the Galactic axions.

The very light axions are restricted by several
laboratory experiments. These include\\
(1) meson decays: $J/\psi\longrightarrow \gamma a,\ 
\upsilon\longrightarrow \gamma a,\ K^+\longrightarrow
\pi^+a, \ \pi^+\longrightarrow e^+\nu_ea$, etc.\\
(2) beam dump experiments: $p(e)+N\longrightarrow a+X,\ 
\ a\longrightarrow \gamma\gamma,\ {\rm or}\ e^+e^-$,\\ 
(3) nuclear deexcitation: $N^*\longrightarrow Na,\ \
a\longrightarrow \gamma\gamma,\ {\rm or}\ e^+e^-$.\\

The above laboratory experiments restrict the decay constant
of the very light axion to
\begin{equation}
F_a > 10^4\ {\rm GeV}.
\end{equation}
However, the laboratory bound is not as strong as the
supernova bound which is discussed in the 
following section.

\section{The Axion Window from Outer Space}

\subsection{Stars}

A stringent lower bound on the axion decay constant 
comes from the study of stellar objects such as Sun, red
giant stars, and supernovae~\cite{axionr}.

In the hot plasma in stars, there exist high energy 
particles: $\gamma, e, p$, and even gluons and quarks
if the interior temperature is high enough. The hot
plasma of the interior of stars, even axions can be
produced. But the axion interaction rate is suppressed
by $F_a$. Bigger stars seemed to give stronger
constraint. For example, the red giant gave a stronger
constraint than Sun, and the supernovae gave
even more stronger constraint than red giant 
stars~\cite{axionr}.
Even before the discovery of SN1987A, there was a
prophetic paper on supernovae by Iwamoto~\cite{iwamoto},
and furthermore most stars
are surveyed by Pantziris and Kang~\cite{kang}.
In this study, the red giant 
constraint was the strongest, giving a
bound $F_a>0.8\times 10^8$~GeV.
Unfortunately, Pantziris and Kang did not give a
useful bound from supernovae, presumably from the
ill-treated pion-nucleon couplings. 
 
The lucky observation of SN1987A gave the strongest
lower limit on $F_a$, $F_a>0.6\times
10^9$~GeV~\cite{supernovae}. 
In the actual calculation, a correct treatment of
the interactions is needed. For example, for the 
pion-nucleon coupling one can use pseudo-vector
derivative couplings and pseudo-scalar couplings.
The reason is that the most conspicuous process of
the axion emmission in supernovae is
\begin{equation}
N+N\longrightarrow N+N+a
\end{equation} 
with a pion exchange diagram. Without a correct treatment 
of this $NN\pi$ coupling, the calculations cannot be
trusted. It was only clarified~\cite{choikang}
after the initial calculations~\cite{supernovae},
but the bound still remains at order $10^9$~GeV.

\subsection{Cosmology}
The axion has a few interesting properties
relevant for cosmology. Since the axion is
created by the spontaneous breaking of a 
global $U(1)$ symmetry, the cosmic strings can be created
in the early universe~\cite{axionr}. The oscillation
of this axionic string can be damped by creating
axions. This was used to restrict the bound on
$F_a$~\cite{Davis}, to less than $10^{11}$~GeV. 
Another interesting object created by axion background
is the axionic domain walls~\cite{DW}, which can be 
problematic in the standard Big Bang cosmology.
However, these problems are avoided in the most
popular cosmological scenario, i.e. inflation with
supergravity Lagrangian. Supersymmetry seems to
be the most popular extension beyond the standard
model, and the inflationary idea seems to be needed
for the flatness and homogeneity problems. In this case,
the gravitino cosmology is inescapable and the reheating
temperature after inflation is better to be less than
$10^9$~GeV~\cite{ekn}. Thus, the problematic axions from
axionic strings and axionic domain walls are all inflated
away.


\begin{figure}
\begin{center}
\begin{picture}(400,460)(-10,0)

\rText(80,400)[][l]{\Large KSVZ}\Text(130,400)[]{\Huge $\Longrightarrow$}
\rText(80,300)[][l]{\Large DFSZ}\Text(130,300)[]{\Huge $\Longrightarrow$}
\LongArrow(180,460)(180,255)\rText(172,380)[][l]{\large$e_3=1$}
\LongArrow(195,320)(195,255)\rText(187,305)[][l]{\small$(u^c,e) \textrm{ unif.}(x=60)$}
\LongArrow(202,280)(202,255)\rText(202,316)[][l]{\small$(u^c,e)(x=1.5)$}
\LongArrow(215,460)(215,255)\rText(207,380)[][l]{\large$e_Q=0$}
\LongArrow(225,460)(225,255)\rText(219,390)[][l]{\large$e_3=-1/3$}
\LongArrow(248,460)(248,255)\rText(243,287)[][l]{\small$(d^c,e)$ unif.}\rText(241,410)[][l]{\large$e_3=2/3 \textrm{ or }e_8=1$}
\LongArrow(256,300)(256,255)\rText(260,308)[][l]{\small nonunif. $(x=1.5)$}
\LongArrow(290,460)(290,255)\rText(282,307)[][l]{\small nonunif. $(x=1)$}\rText(282,380)[][l]{\large $(m,m)$}

\LinAxis(306,0)(30,0)(6.4,1,5,0,1.5)
\LinAxis(306,250)(30,250)(6.4,1,0,0,1.5)
\rText(89.5,-25)[][l]{\Large$10^{-16}$}
\rText(175,-25)[][l]{\Large$10^{-18}$}
\rText(264,-25)[][l]{\Large$10^{-20}$}

\LogAxis(306,0)(306,250)(0.8,5,2,1.5)
\LogAxis(30,0)(30,250)(0.8,0,2,1.5)
\rText(318,0)[][l]{\Large$3\times10^{-6}$}
\rText(318,210)[][l]{\Large$10^{-5}$}
\rText(335,120)[][l]{\Large$m_a[eV]$}
\rText(-20,120)[][l]{\Large$F_a[GeV]$}
\Line(30,120)(23,120)
\rText(13,120)[][l]{$10^{12}$}

\DashLine(267,0)(267,250){4}
\rText(275,120)[][l]{\large CARRAK II}
\Text(261,210)[]{\small$\Longleftarrow$}
\Text(261,200)[]{\small$\Longleftarrow$}
\Text(261,190)[]{\small$\Longleftarrow$}
\Text(261,40)[]{\small$\Longleftarrow$}
\Text(261,30)[]{\small$\Longleftarrow$}
\Text(261,20)[]{\small$\Longleftarrow$}
\Line(253,0)(253,90)\rText(245,115)[][l]{\large LLNL}\Line(253,140)(253,250)
\rText(255,115)[][l]{\large upgrade}
\Text(240,50)[]{\Large$\Longleftarrow$}
\Text(240,20)[]{\Large$\Longleftarrow$}
\Line(223,0)(223,90)\rText(220,115)[][l]{\large LLNL}\Line(223,140)(223,190)
\rText(230,115)[][l]{\large now}
\Text(210,50)[]{\Large$\Longleftarrow$}
\Text(210,20)[]{\Large$\Longleftarrow$}

\Text(20,12)[]{\Large$\Longrightarrow$}
\rText(5,0)[r][l]{\large$2001$ MIT-F-B $\cdots$}
\Text(20,80)[]{\Large$\Longrightarrow$}
\rText(5,84)[r][l]{\large RB }
\Text(20,135)[]{\Large$\Longrightarrow$}
\rText(5,145)[r][l]{\large LLNL }
\Text(20,207)[]{\Large$\Longrightarrow$}
\rText(5,225)[r][l]{\large CARRAK I}
\rText(178,-60)[][u]{\Large$c_{a\gamma\gamma}^2 F_a^2\times factor$}

\Line(107,201)(30,201)
\Line(107,195)(107,201)
\Line(100,195)(107,195)
\Line(100,188)(100,195)
\Line(60,188)(100,188)
\Line(60,178)(60,188)
\Line(78,178)(60,178)
\Line(78,155)(78,178)
\Line(82,155)(78,155)
\Line(82,151)(82,155)
\Line(86,151)(82,151)
\Line(86,147)(86,151)
\Line(90,147)(86,147)
\Line(90,85)(90,147)
\Line(120,85)(90,85)
\Line(120,65)(120,85)
\Line(30,65)(120,65)

\GBox(30,205)(248,210){0.8}
\GBox(30,160)(124,163){0}
\GBox(30,140)(116,144){0}
\GBox(30,120)(120,140){0}
\GBox(30,100)(140,120){0}
\GBox(30,0)(190,28){0.5}

\end{picture}
\vskip 3.2cm
\caption{The status of the axion search experiments
of Rochester-Brookhaven, Florida, LLNL, Kyoto, and
MIT-Fermilab experiments.}
\end{center}
\end{figure}
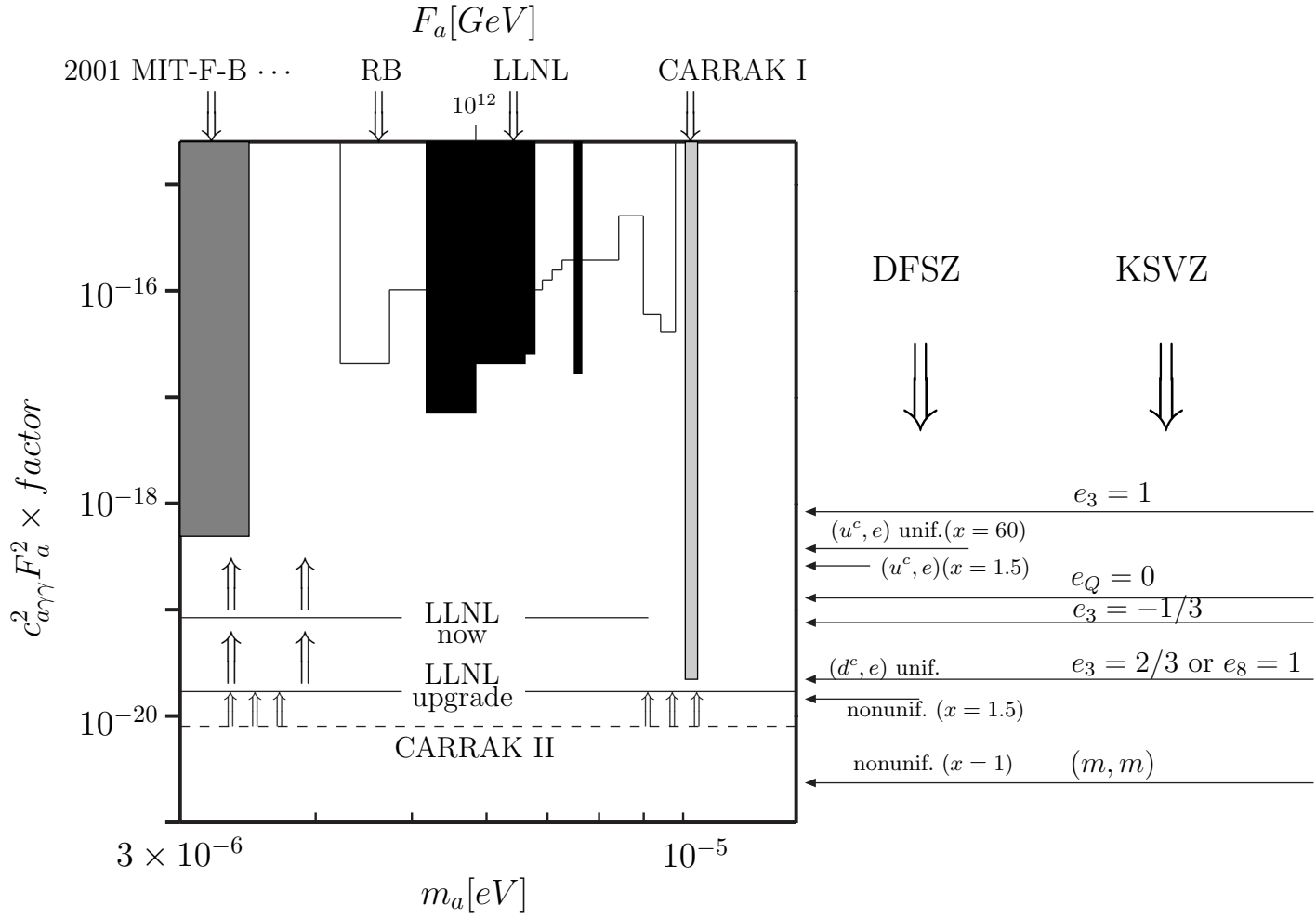

After the inflation, the value of the classical axion field
in our homogeneous patch is of order $F_a$. For $\langle
a\rangle\ll F_a$, the probability is so small. Therefore,
in general this initial value of the classical axion field
is supposed to be of order $F_a, \langle a\rangle_{RH}\equiv
F_1\sim O(F_a)$. This initial value of $\langle a\rangle$
stays there for a long time since the axion potential
is extremely flat as shown in Fig. 1. The axion vacuum
starts to oscillate when the Hubble expansion rate($1/H$) 
became small enough to be comparable as the classical axion 
oscillation rate($1/m_a$), which happens when the temperature
of the universe falls to around $T_1\simeq1$~GeV~\cite{pww}. 
Below $T_1$ the oscillating classical axion field carries the
energy density of order $m_a^2\langle a\rangle^2$. As the
universe expands, the amplitude of the axion field shrinks,
which is the result of the conserving axion number density
$m_a\langle a\rangle^2$ in the comoving volume. Taking into
account of this evolving axion universe, one 
obtains~\cite{pww,turner,axionr},
\begin{equation}\label{density}
\rho_a(T_\gamma)=m_a(T_\gamma)n_a(T_\gamma)
=2.5\frac{F_a}{M_p}\frac{m_aF_a}{T_1}T_\gamma^3\left(
\frac{\langle a\rangle(T_1)}{F_a}\right)^3.
\end{equation}
In Eq.~(\ref{density}), the factors except $\frac{F_a}{M_p}$
gives roughly $10^7\rho_c$ where $\rho_c$ is the current
critical energy density. Therefore, not to overclose the 
universe by the oscillating axion field, $F_a$ should be
less than $10^{12}$ GeV. 

Summarizing the astro- and cosmological- bounds, we arrive at
an axion window still open to be observed,
\begin{equation}
10^9~{\rm GeV}< F_a< 10^{12}~{\rm GeV}.
\end{equation}
Note, however, that if the axions constitute only 25\% of
the critical energy density then $F_a$ should be $\frac14$
of the above estimate. However, we will take $10^{12}$~GeV
as a generic number since there can exist other 
uncertainties such as the initial value of $\langle a\rangle$,
the domain wall number, etc.

\subsection{Galactic axion search}
If the axion is the CDM component of the universe, it is
pointed out that they can be detected~\cite{cavity}. Even
though the interaction is inversely proportional to $F_a$,
the enormous number density cancels this effect, and the
galactic axions can be detected. Indeed, there have been 
several experiments since late 1980's~\cite{asearch,akyoto}.
The cavity~\cite{asearch} and Rydberg atom~\cite{akyoto}
searches utilize the axion-$F\tilde F(photon)$ coupling.
Even though the $\theta F\tilde F(photon)$ is absent since
$\theta$ is a parameter, the axion-$F\tilde F$ is a
physically observable operator~\cite{cavity}. Sikivie
considered this axion-$F\tilde F$ coupling in the
presence of strong gradient of magnetic field of
order 10 Tesla.

In Fig. 4, we summarize the current status of the axion search
experiments. There are two points to be clarified here.
One is that there are many very light axion models, in KSVZ
type and in DFSZ type. The definition of the notation is
given in Kim~\cite{kim98}. For example, an arbitrary choice
of $e_3=1$ cannot fulfil the unification of couplings,
and hence should not be considered toward unification
of elementary particle forces~\cite{kim03}. 
Another point is that these estimates are based on
closing the universe by axions. If it is required that
only 25\% of the critical energy is closed by the axions,
the constraint on the axion density is
strengthened by a factor of 4.

It is noted that still there is a lot of room to allow
the very light axion as CDM.

\section{The Quintaxion (Quintessential Axion)}
Typa 1a supernova data and WMAP data hints that
73\% of the energy of the universe is homogeneous
dark energy~\cite{WMAP}, and hence does not contribute
the galaxy formation unlike the CDM candidates.
In this section, we try to introduce an axionlike 
particle for the dark energy, which is called a
quintessential axion or quintaxion. As this name
suggest quintaxion is a kind of pseudo Goldstone
boson whose potential arises from instanton 
effects. This kind of extremely light ($\sim 10^{-33}$~eV)
pseudo-Goldstone boson was suggested sometime ago~\cite{stebins}.
The attempt to incoorporate it in superstring or M-theory
was also tried~\cite{stringax}. There have been another
kind of scalars whose present vacuum energy 
is required to be $\sim (0.003~{\rm eV})^4$, which are
generally called quintessence~\cite{quint}.

The condition for quintaxion is $\omega=p/\rho<-
\frac13$,\footnote{But the
recent WMAP data~\cite{WMAP} suggests 
a stronger $\omega<-0.8$.}
For a pseudo-Goldstone boson $a_q=\theta f$ with the potential
$U(\theta)$ and the decay constant $f$, $\omega$ is given as
\begin{equation}
\omega=\frac{-6f^2+M_p^2|U^\prime|^2}{6f^2+M_p^2|U^\prime|^2}.
\end{equation}
With $f$ near the Planck scale a quintaxion is
natuarally realized~\cite{stebins}. For superstring axions,
the decay constants are expected to be of order Planck scale
but the model-independent axion decay constant is of
order $10^{16}$~GeV~\cite{harmful}. Except this special
case, we generally anticipate that string axions such
as the model-dependent ones would have a Planck scale
axion decay constants. So, even though we present 
formulae for the model-independent axion for a concrete
presentation, this caviat is implied. For this quintessential
axion to work, we require that the potential is extremely
flat with the height of order $(0.003~{\rm eV})^4$ with
a Planck scale order decay constant.

From our discussion on the massless quark in Sec. 1,
it is very tempting to use some kind of almost massless quarks
for the extremely flat quintessential potential.
Since a hidden sector is needed for supersymmetry breaking,
we will attempt the hidden-sector quarks(h-quarks) as
fulfilling this purpose. Before presenting a quintaxion
along this line, let me comment a few words on the
superstring axion.

Among several possibilities of the QCD axion given in
Table 1, the superstring axion seems to be the 
most attractive realization. The axion is definitely
the most attractive solution of the strong CP problem,
because it arises by automatically solving the strong CP 
problem. It relies only on the $aF\tilde F$ coupling.
On the other hand, the other solutions should massage
the theory much, by postulating specific forms of
couplings. For example, $m_u=0$ possibility is not
that simple theoretically. It requires that 
Det.~$M_{\rm up}=0$, which is a highly nontrivial 
requirement for the nine complex entries of $M_{\rm up}$. 
Similarly, superstring axions is the most attractive 
since all string compactifications render a plenty
of axions. For the other axions, one massage the
theory more compared to the string axions. In this regards,
if the model-independent axion were acceptable as the
QCD axion, then it might have given a very reliable
axion candidate.

However, the model-independent axion failed miserably
because of a too large axion decay constant~\cite{harmful}.
So, as the first principle the string axion as the QCD
axion failed.  One must remove this model-independent axion 
at high energy. Indeed, it is possible to do so, if there
exists the anomalous $U(1)$ gauge symmetry so that
the $U(1)$ gauge boson eats up the model-independent axion
as its longitudinal degree~\cite{sen}. 
But, then there must survive a global symmetry which
must be broken at the intermediate scale~\cite{kim88}.
In another context, some discrete symmetries were
invoked to obtain an approximate global symmetry, approximate
in the sense that the global symmetry breaking operators
are allowed at dimension 13 and higher~\cite{lshafi}.
In these cases, string models can allow a QCD axion.

Here, we are interested in obtaining the QCD axion also 
together with a quintaxion in string inspired 
models~\cite{quintaxion}. We are arrived at this kind
of scenario from old solutions of the $\mu$ problem~\cite{mu}
through the composite QCD axion~\cite{chun}.
The superstring axions couples to both the QCD anomaly and
the hidden sector anomaly. In this case, we must consider
two $\theta$ angles, $\theta_{QCD}$ and $\theta_h$. To
have the vacuum at $\theta_{QCD}=\theta_h=0$, we need two
axions. Here, we want to introduce two such axions as
the composite one from the hidden-sector squark condensation
and a superstring axion. In this case, there exists a
problem that an axion with a higher potential
corresponds to the smaller axion decay constant and a
shallower potential corresponds to the larger decay 
constant~\cite{stringax}. Since the hidden sector is expected 
to be at $\Lambda_h\sim 10^{13}$~GeV, the QCD axion generally
corresponds to the Planck mass decay constant, grossly 
violating energy bound~\cite{pww}. This correspondence
is shown in Fig. 5 as real lines. We have to change this
behaviour to introduce a reasonable QCD axion. 


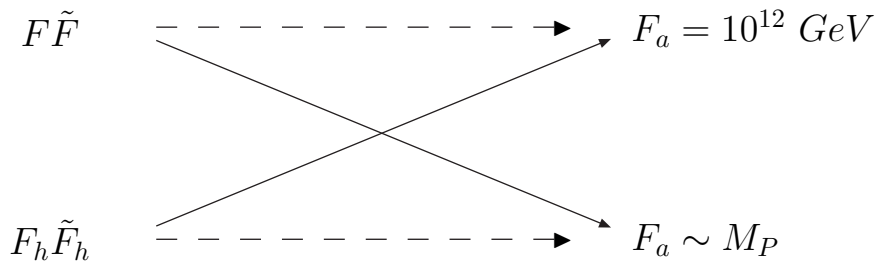
\begin{figure}
\begin{center}
\begin{picture}(300,150)(10,0)
\DashLine(50,130)(200,130){8}\Text(200,130)[l]{$\blacktriangleright$}
\DashLine(50,50)(200,50){8}\Text(200,50)[l]{$\blacktriangleright$}
\LongArrow(50,55)(220,125) \LongArrow(50,125)(220,55)
\Text(10,130)[c]{\Large$F\tilde{F}$}
\Text(10,50)[c]{\Large $F_h\tilde{F}_h$}
\Text(230,130)[l]{\Large$F_a=10^{12}\ GeV$}
\Text(230,50)[l]{\Large$F_a \sim M_P$}
\end{picture}
\vskip -1.2cm
\caption{One naively anticipates that the QCD axion corresponds to
a Planck scale decay constant. Our objective is that the QCD axion
should corresponds to $F_a\simeq 10^{12}$~GeV as shown with the
dashed arrows.}
\end{center}
\end{figure}

It seems that the string axion problem is a real problem.
However, if the hidden-sector instanton potential is
made shallower than the QCD axion potential, the QCD axion
can be saved, as shown in Fig. 5 as dashed lines. 
In this regards, we adopt the idea of the
almost massless h-quarks. But if the h-quark mass is
exactly zero, then there is no quintessence. So we
must render the h-quarks tiny masses.
As we have seen in Sec. 1, the current quark masses come in the
determination of the instanton potential. For the hidden-sector 
vacuum, we assume almost massless hidden-sector quarks.
Also hidden sector gluino mass ($\sim 100$~GeV) is much 
smaller than the hidden sector scale $\Lambda_h\sim 10^{13}$~GeV.

A solution of the $\mu$ problem is assuming a Peccei-Quinn
symmetry,\footnote{Supersymmetry breaking in supergravity can
generate a $\mu$ term~\cite{giudice}. But here also, one
must assume the absence of $H_1H_2$ in one way or another.} 
forbidding the $H_1H_2$ term and $Q_i\bar Q_j$ term
in the superpotential. The Peccei-Quinn symmetry, however,
allows~\cite{chun}
\begin{equation}\label{hhqq}
W_\mu=\frac{c}{M_p}H_1H_2Q_i\bar Q_j
\end{equation}
where $c$ is an order 1 number.
If the h-squarks $Q_i{\bar Q_j}$ condenses
at the intermediate scale we obtain a $\mu$ term at the
electroweak scale order. This is one side of a coin.
The other side of coin is that if the Higgs fields
develops vacuum expectation values at the electroweak
scale, $v\simeq 248$~GeV, then the h-quarks obtain
tiny masses of order
\begin{equation}\label{mq}
m_{Q}\simeq 0.64\times 10^{-14}\sin 2\beta~[\rm GeV]
\end{equation}
where $\beta=\langle H_2^0\rangle/\langle H_1^0\rangle$.

For small instantons, the height of the potential is estimated
to be, for $n$ h-quark flavors and $N$ number of colors,
\begin{equation}
\lambda_h^4\simeq m_Q^n m_{\tilde G}^N\Lambda_h^{4-n-N}.
\end{equation}
Thus, the hidden sector potential can be extremely flat,
contrary to a naive expectation.
It is because the h-quark mass and h-gluino mass
are much smaller than the hidden sector scale $\Lambda_h$.
Anyway, even though it looks weird, the hidden sector
axion potential is much smaller than the QCD axion 
potential, realizing the dashed line correspondence
of Fig. 5~\cite{quintaxion}. It has been much pronounced 
by almost the massless quark.

Summarizing, we needed two axions which are provided by
(1) superstring axions, either the model-independent axion 
or model-dependent axions which becomes the
quintaxion, providing the dark energy and (2) the axion through 
introducing a Peccei-Quinn symmetry needed for solving the
supergravity $\mu$ problem. The h-squark condensation
introduces another axion which becomes the QCD
axion, providing the CDM in the universe.

\section{Conclusion}

In this talk, I reviewed the current status of the
strong CP problem and axion. Regarding the QCD axion,
we showed also the current status of the axion
search bound, assuming that the QCD axion is the
dominant component of the CDM. Furthermore, we
speculated that the axionlike particles present
in superstring models can be candidates for the
source of the dark energy in the universe.

\vskip 0.5cm
\noindent{\bf Acknowledgments}

This work is supported in part by the KOSEF
ABRL Grant to Particle Theory Research Group
of SNU, the BK21 program of Ministry of Education,
and Korea Research Foundation Grant No. 
KRF-PBRG-2002-070-C00022.

\end{document}